\documentclass[english, aps, prb, twocolumn, noshowkeyws, noshowpacs, superscriptaddress, reprint]{revtex4-1}
\usepackage{color}
\usepackage{textcomp}
\usepackage{graphicx}
\usepackage{hyperref}

\begin{document}

\title{A portable MBE system for in situ X-Ray investigations at synchrotron beamlines}

\author{T. Slobodskyy}
 \email{Taras.Slobodskyy@physik.uni-hamburg.de}
 \affiliation{Institute for Synchrotron Radiation, Karlsruhe Institute of Technology - 76344 Eggenstein-Leopoldshafen, Germany}
 \affiliation{Institut f\"ur Angewandte Physik und Zentrum f\"ur Mikrostrukturforschung,\\Jungiusstra\ss e 11, D-20355 Hamburg, Germany}
\author{P. Schroth}
\author{D. Grigoriev}
\author{A. A. Minkevich}
 \affiliation{Institute for Synchrotron Radiation, Karlsruhe Institute of Technology - 76344 Eggenstein-Leopoldshafen, Germany}
\author{D. Z. Hu}
\author{D. M. Schaadt}
 \affiliation{Institute for Applied Physics/DFG-Center for Functional Nanostructures,Karlsruhe Institute of Technology (KIT), Karlsruhe, Germany}
 \affiliation{Institute for Energy Research and Physical Technologies, TechnicalUniversity Clausthal, Am Stollen 19B, 38640 Goslar, Germany}
\author{T. Baumbach}
 \affiliation{Institute for Synchrotron Radiation, Karlsruhe Institute of Technology - 76344 Eggenstein-Leopoldshafen, Germany}

\begin{abstract}

A portable synchrotron MBE system is designed and applied for \textit{in situ} investigations. The growth chamber is equipped with all the standard MBE components such as effusion cells with shutters, main shutter, cooling shroud, manipulator, RHEED setup and pressure gauges. The characteristic feature of the system is the beryllium windows which are used for \textit{in situ} x-ray measurements. An UHV sample transfer case allows in-vacuo transfer of samples prepared elsewhere. We describe the system design and demonstrate it's performance by investigating the annealing process of buried InGaAs self organized quantum dots.

\end{abstract}


\maketitle

\section{Introduction}

The molecular Beam Expitaxy (MBE) is a versatile technique allowing precise control over the deposition environment in a wide range of growth conditions \cite{arthur_molecular_2002}. It is often used for deposition of self organized nanostructures on semiconductor surfaces \cite{joyce_effect_2000}. Improvement of the properties of the nanostructures beyond current state of the art requires understanding of dynamical processes during self organization. Strain induced interaction between the nanoobjects \cite{riotte_lateral_2010} as well as change in the strain state of the nanostructures during deposition and annealing \cite{hu_morphology_2010} are of specific interest for the scientific community. 

The strain state of the nanostructures can be accessed using X-ray diffraction \cite{metzger_x-ray_2005}. Laboratory x-ray diffractometers are routinely used for strain investigation. The diffractometers, however, hardly permit integration of such a demanding equipment as a growth chamber for MBE into the measurement setup. Beside this, the intensity of the x-ray beam produced using laboratory setup is not sufficient to characterize dynamical processes at nanoscale with sufficient time resolution.

High brilliance synchrotron radiation provides a way to overcome these difficulties \cite{fong_x-ray_2010}. The high brilliance of the x-ray beam allows for fast scan repetition and highly parallel beam permits reasonable size of the deposition equipment. Therefore we decided to use synchrotron radiation for investigation of dynamical processes during MBE deposition.

In the past there have been several approaches to combine UHV equipment
and MBE systems with synchrotron diffractometers \cite{albrecht_six-circle_1999,barbier_new_1999,bernard_ultrahigh_1999,couet_compact_2008,fuoss_apparatus_1984,jenichen_combined_2003,stankov_ultrahigh_2008,tajiri_sample_2004,vlieg_ultrahigh-vacuum_1987,maree_system_1987}.
In principle, these concepts can be separated to two general philosophies.

The first approach is to use a large stationary UHV system in
which a diffractometer is integrated. This is usually advantageous
in terms of vacuum quality and growth conditions, because the overall
weight of the MBE components is not limited by additional constraints such as mobility of the system.
This holds as well for the vacuum system, and supplementary surface
analysis equipment. However, a complete beamline has to be dedicated
to this single purpose in this case. 

The second approach is to utilize a small
lightweight UHV chamber that is installed at a suitable diffractometer
at a synchrotron beamline on demand. The latter approach often lacks
adequate control over the growth parameters and scanning motions of
the diffractometer. Yet, it gives versatile possibilities in choosing
the appropriate beamline suitable for the particular X-ray experiment.
We chose the latter philosophy in order, to have a standalone MBE
system which, nevertheless is portable and suitable to be mounted
on a diffractometer at a synchrotron beamline of our choice. 

In this paper, we describe a portable UHV system which can be used at various beamlines but without compromising on quality of epitaxial nanostructures.
The system is developed for in situ investigations of growth and evolution of epitaxial semiconductor nanostructures.

\section{System layout and technical design}

\begin{figure*}
\includegraphics[width=12cm]{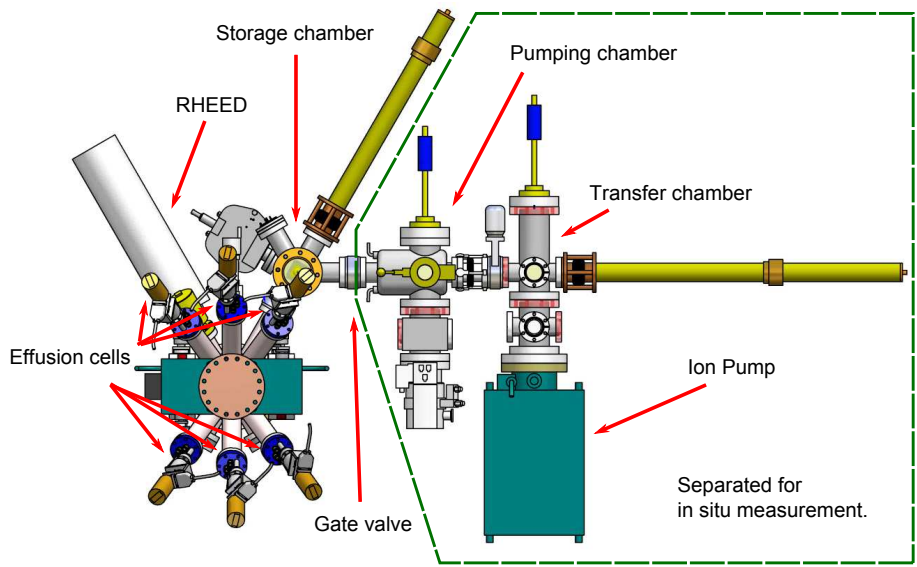}

\caption{Scheme of the portable MBE system. The green box indicates the transfer
case which is separated during X-ray experiments. \label{fig:Scheme}}

\end{figure*}

It was our aim to design a portable MBE system which should be able
to transport and process samples grown at the MBE clusters of our
collaboration partners under UHV conditions. In addition growth of nanostructures
should be possible within this chamber when it is mounted on the diffractometer
and the X-ray experiment is running. The system should offer a possibility
for X-rays to enter, hit the sample and leave the chamber in a widest
angular range possible, to afford X-ray investigations. Let us now discuss how these aims were realized.

The portable MBE system comprises three main parts, a unit for sample
UHV transfer and loading called Transfer Case (TC,) a unit for sample
storage called Storage Chamber (SC), and a chamber for MBE growth
called Growth Chamber (GC), featuring two beryllium windows for the
X-rays as it depicted in the Fig. \ref{fig:Scheme}. To make this system as versatile and mobile as possible, these
three parts are located on a steel frame supported by wheels. 

The
SC and GC can be decoupled from the TC and relocated from the frame
onto the diffractometer for X-ray measurements. To maintain the vacuum
during each step of the experiment, each separate chamber is individually
pumped. 

Despite the fact, that specialized equipment and parts could yield a considerable decrease in the weight of the UHV system, we resort to standard CF components where it is possible. Therefore, the UHV system comprising of GC, SC and TC has a total weight of 350kg. Standard components, however, reduce delays in the case of component failure and decrease operational costs. Consider that the combined GC and SC weight is about 155 kg only. 

Experimental layouts have been realized at the NANO beamline at ANKA - the synchrotron facility at the Karlsruhe Institute of Technology (KIT) and at the
surface diffraction beamline ID03 at the European Synchrotron Radiation Facility (ESRF), where the first \textit{in situ}
measurements have been performed.

\subsection{Growth chamber}

The GC is used for MBE-growth. The inner diameter of the GC is 250~mm. A cooling shroud is installed into a CF150 flange, and a CF100
flange is used for the ion pump which is equipped with a titanium
sublimator. Two CF100 flanges are dedicated to standard 90mm beryllium
X-ray windows. CF40 ports for sample transfer, Reflection High Energy
Electron Diffraction (RHEED) gun and RHEED screen as well as two view
port flanges are present. Six CF40 flanges are pointing to the sample
surface and can be used to install effusion cells. 

In the present configuration, three of the effusion cells are used for growth
of InGaAs compounds, pointing at the sample surface normal at an angle
of 30\textdegree{}. The effusion cells for indium and gallium are
high temperature dual-filament cells whose crucibles and tips can
be separately heated. The effusion cell for arsenic is a low temperature
cell with one heating zone. The cells are equipped with fast shutters
in front of the cell\textquoteright{}s orifices. 

The GC is equipped with a main shutter which can be positioned in front of the sample
to protect the surface during the effusion cells adjustments. The
precise positioning of the shutter can also be used for low resolution
shadow epitaxy when it is required \cite{PhysRevB.70.155328}.

A manipulator is responsible for sample positioning featuring translation and endless-rotation is motorized by brushless DC servo-motors "Faulhaber".
A pneumatically actuated stabilization stamp can fix the sample position
during X-ray measurements. Using the manipulator, the sample can be
positioned such that RHEED and X-ray measurements can be performed
at the same time. In this configuration the sample is located in the
center between the beryllium window ports. Tungsten heating wires
provide sample heating. The sample heater supports heating up to T=900\textdegree{}C.
The temperature at the sample is measured by a thermocouple located
in the substrate heater near to the substrate. 

The sample is mounted such, that it's surface normal points in horizontal direction. This is necessary to avoid a decrease of the
scattered intensity due to polarization effects for high scattering
angles in Grazing Incidence Diffraction (GID) geometry, keeping in mind, that synchrotron radiation is polarized in
the horizontal plane. We designed a special sample-holder to keep
the sample in its vertical position, not blocking RHEED or X-ray beams.

\begin{figure}
\label{Sample on manipulator}{\includegraphics[width=8.5cm]{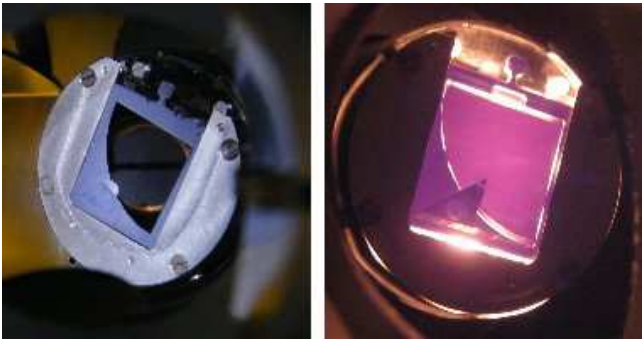}}

\caption{Left: View of a sample mounted on the manipulator. Right: Heating uniformity of the sample at high temperature.}
\label{fig:manip}
\end{figure}

The standard samples used in our system have the size of a quarter of a 2" wafer. This provides
a sufficiently large opening at the back of the sample holder to achieve a homogeneous
temperature distribution over the sample. A second advantage of the
sample size is the fact, that the large beam-footprint resulting of
the grazing incidence geometry will completely be projected on the
sample surface. Thus unfavorable edge effects of the scattered and
reflected beam can be avoided \cite{couet_compact_2008}. 
Figure \ref{fig:manip} shows a view of a sample mounted on the manipulator. The sample is seen at different temperatures and from different view ports.

\begin{figure*}
{\includegraphics[width=16cm]{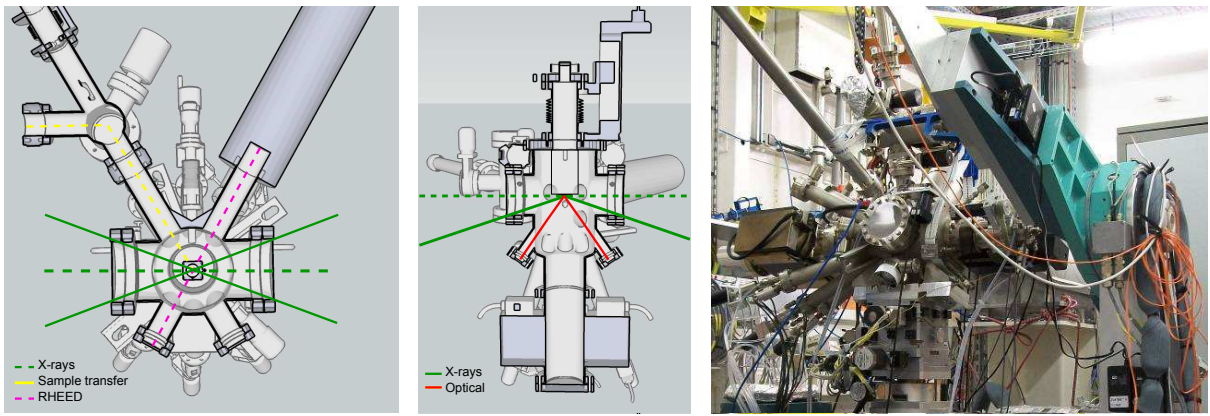}}

\caption{a) vertical cut through growth chamber. The sample position is in
the crossing points of the lines. Pink: RHEED, Green: X-Ray, Yellow:
sample transfer. b) Horizontal cut through growth chamber. Green:
X-Ray, Orange: Optical, Blue: Material from effusion cells. c) Growth
chamber installed at ID03 vertical diffractometer, ESRF. Left: Effusion
cells, Ion pump and TSP. Center: Beryllium window pointing to the
detector with flight tube. Right: diffractometer detector arm. \label{fig:geometry}}
\end{figure*}

Three view ports are installed to visually check the sample position
for transfer. Additionally, the view ports are used for a LASER pre-alignment
of the sample surface during X-ray measurements. The two beryllium-windows,
located on the opposite sides of the deposition chamber (see Fig. \ref{fig:geometry}), allow \textit{in situ} X-ray measurements. Motorized shutters in front of the
beryllium-windows prevent the contamination of the beryllium-windows
during growth when no X-ray experiments are performed. The same shutters
can be used as in-vacuum X-ray beam stop to decrease the effect of
parasitic scattering on the exit Be-windows when it is needed.

A cold-cathode pressure gauge "Pfeiffer" provides information about
the vacuum conditions in each of the chambers. A 100
l/s Ion pump "Gamma Vacuum" combined with a titanium sublimation pump achieve vacuum
conditions with a base pressure in the lower $10^{-11}$~mbar range.
The cells and the manipulator are water-cooled. Chilled water or liquid
nitrogen can be supplied to the cryoshroud of the GC to improve vacuum
conditions even further.

\begin{figure}
\label{Sample on manipulator}{\includegraphics[width=8.5cm]{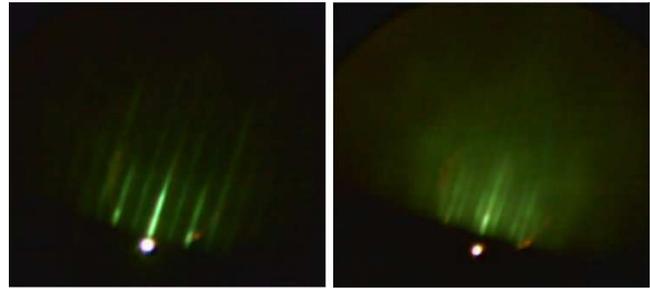}}

\caption{(2x4) RHEED reconstruction of a GaAs surface after oxide desorption observed during an \textit{in situ} experiment.}
\label{fig:rheed}
\end{figure}

A 30~keV reflection high energy electron diffraction (RHEED)
gun "SPECS" is operated for monitoring the growth process. The RHEED pattern
is projected to an opposing fluorescence screen on a CF40 flange
(pink line Fig. \ref{fig:geometry}, left). Besides RHEED is an effective
tool to investigate the surface quality and growth speed, it is commonly
used to monitor the growth of nanostructures, especially to determine
the 2D-3D transition during quantum dot growth, for example. Utilizing the motorized manipulator for sample rotation various
surface reconstructions can be investigated. Two RHEED patterns of a (2x4) GaAs surface are shown on the Fig. \ref{fig:rheed}. 

\subsection{Storage chamber}

The storage chamber (SC) is directly connected to the GC. It is equipped
with a magnetically operated storage cassette. This cassette can hold
up to four samples. Using the transfer rod connected to the SC the
samples are transferred into the GC and placed onto the manipulator
for growth and X-ray investigation. During growth, the two chambers
are isolated via a gate valve and the SC is pumped separately by a 10s ion pump "Gamma Vacuum". 

\subsection{Transfer case}

This part of the portable MBE system is used for UHV sample transfer
to and from other MBE clusters and beamlines, like the III/V MBE cluster
at the Center for Functional Nanostructures (CFN) at KIT to the dedicated
synchrotron beamlines at ANKA or ESRF respectively. 

As well as the
GC and SC, the TC consists of two individually pumped UHV parts which
can be separated via a gate valve. The pumping chamber is used as
a load-lock for inserting samples. It can be vented with dry nitrogen.
Vacuum is achieved by a turbo molecular pump (TMP) "Pfeiffer" connected
to a scroll pump. The pressure is measured by a cold-cathode
wide range gauge "ATMION". An infra-red heating lamp can be used to degas the
loaded samples. 

After degassing in the pumping chamber, the samples
can be transferred into the SC or into the second chamber of the TC where  up to 10 samples can
be stored in the storage cassette. The chamber holding the
storage cassette can be isolated from the pumping chamber by a CF40 gate
valve, and is pumped separately by a 150~l/s ion pump "Gamma Vacuum"
and a TSP to achieve best possible vacuum during the sample transfer
and storage. The pressure is monitored by a Bayard-Alpert hot cathode
pressure gauge.

\subsection{Integration and MBE control software}

Due to hazardous radiation inside of the experimental hutch during synchrotron \textit{in situ} experiments the MBE system should be fully automated and remotely controlled. All of the MBE system components should work simultaneously and synchronously. They also should interact with the beamline control software. Consequently already available standard solutions could not be easily adapted for our needs.

For the control purpose we have written a home made multithreading MBE control software, which is scriptable, object oriented, and cross-platform. Bidirectional communication with SPEC and Tango is already realized in the software. Extension to other interfaces should not be difficult through python written drivers.

The effusion cells shutters, main shutter, beryllium window shutters, manipulator translation and rotation are controlled using precise position sensitive servo motors. This allows for precise adjustment of the experimental conditions without external intervention. The cells temperatures are controlled by digital controllers "Eurotherm". And pressure inside of all of the UHV components is monitored using cold cathode gauges as it is mentioned before.

The MBE control software is written in Python. Some of the advantages of this programming language are the easy syntax, cross platform nature and rich scientific libraries which facilitate application of this interpreted programming language in various areas of scientific and automation tasks. The graphical interface is realized using wxWidgets but can be easily transfered to Qt or GTK+.

The control software makes use of the internal python interpreter through Python execution calls. Functions specific for our automation tasks are predefined in the execution environment. The interpreter provides us with a complete, mature, and verified script execution environment. The Python syntax of the scripts makes it easy to learn to operate the software. The interpreter also takes care about the syntax, logic, threads, and interfaces.

The equipment is controlled using standard Ethernet topology through TCP/IP socket communication. For some control devices Ethernet to RS232 converters are used for interfacing. The control software is usually executed on a standard notebook computer without special adaptations. This makes the setup very flexible and failure prone, since the control computer can be rapidly replaced.

Multiple scripts can be executed, paused and interrupted simultaneously. Execution environment of an individual script can be accessed from the command line.

During operation of the MBE software several threads are normally executed simultaneously performing such tasks as communication to the devices, recording of the growth and vacuum conditions, communication with beamline software and execution of user growth scripts.

\subsection{RHEED software}

Reflection high energy electron diffraction is routinely used for growth monitoring. The CF40 RHEED view port of our growth chamber is located on the distance of 300~mm from the center of the substrate. This provides sufficient angular range for electron diffraction pattern observation at 24~keV. The electron gun provides sufficiently small focal spot compensating for degradation of angular resolution of the setup due to small sample-screen distance.

We use also a home made software for observation of RHEED patterns. The RHEED software can communicate with the MBE growth software. This allows us to synchronize sample rotation and growth stages with surface morphology observed through RHEED patterns. The data can be also streamed through the network to another computer for remote operation during \textit{in situ} experiments.

The RHEED analysis software is written in python programming language and uses standard windows DirectShow video capture interface. In this manner various video-capture devices (e.g. a USB webcam) can be connected simultaneously and can be accessed on demand. The software was tested with few consumer webcams providing frame rates of up to 35 frames per second ($\approx$30~ms time resolution) and image resolutions of up to 1600x1200 pixels.

Flexible scripting allows the analysis software to be used for simultaneous pattern profiling. Currently implemented scripts include region of interest intensity monitoring for growth rate determination, linear profile evolution for lattice constant measurements, and frame recording for later analysis. The collected data is normally recorded to the hard drive in real time using common formats (eg. ASCII, PNG, TIFF, JPG).

Sample images of RHEED reconstructions collected using the software are shown in the Fig. \ref{fig:rheed}.

\section{X-ray experiment layout}

Crucial for X-ray measurements are the Beryllium windows, fabricated
by Brush Wellman, Electrofusion Products. The two Be-windows are mounted
on CF100 standard flanges which allow for a vacuum in the lower $10^{-11}$~mBar range. The Beryllium foil used in the windows has a thickness
of 300~\textmu{}m. 

Both windows are installed at opposing sites at
the growth chamber, facing the manipulator (see Fig. \ref{fig:geometry}).
Thus, it is possible to apply surface sensitive X-ray Diffraction
methods like GID and Grazing Incidence Small Angle X-ray Scattering (GISAXS) within an angular range of $\pm$ 18\textdegree{}
with respect to the incident beam (Green lines). Furthermore, Optical
reflectivity measurements can be applied to investigate the morphology
of the grown layers. As it can be seen, the sample position (Red)
is such that the surface normal points into horizontal direction.

For X-ray measurements, the GC together with SC are detached from
the transfer case and mounted on the diffractometer at a suitable
synchrotron beamline (see Fig. \ref{fig:geometry}, right). Due to
the 155~kg weight of the growth chamber a heavy-duty diffractometer
(e.g. ID03 at ESRF or NANO at ANKA) is required. 

If the diffractometer
does not support heavy weight e.g. $m_{GC}=155$~kg, a weight-compensation
device using a ceiling mounted crane in the experimental hutch can
be employed to partially release weight-load from the diffractometer.
This, however, should be used with care since it is difficult to balance
the torque and the moments of inertia at the same time, especially
while operating a multi-axis positioning system "Huber tower".

We used a special adaptor plate to mount the growth chamber on the
diffractometer at ID03 for our first \textit{in situ} experiment. This adaptor
plate is used to incline the chamber with respect to the horizontal plane
to an angle of 13\textdegree{}. Hence, the goniometer movement is closer to the {[}220{]} GaAs Bragg reflection and the accessible angular range
$2\theta$ is enlarged from 18\textdegree{} to 31\textdegree{}.

After mounting the chamber on the diffractometer the sample alignment
is done using the X-ray beam. Once the proper measurements conditions
are reached for one of the samples a laser pointer is installed so
that the laser beam comes through one of the view windows and hits
the sample. In this configuration the reflected laser beam is projected
on one of the walls of the experimental hutch. The position of the
reflected beam is labeled on the wall. This position is than used
for all the following samples as a point of reference for alignment.
This procedure speeds up the alignment procedure during \textit{in situ} measurements. 

Choosing the appropriate energy such that reflected and diffracted
beam make an angle less than 31\textdegree{}, GID and GISAXS measurements
can be performed simultaneously. In our case {[}220{]} GID reflex
of InGaAs QDs could be reached using 13~keV photon energy. Detector
frames during simultaneous GID and GISAXS measurements are shown below
in Fig. \ref{fig:gid gisaxs}.

\begin{figure}

\label{Geometry}{\includegraphics[width=8.5cm]{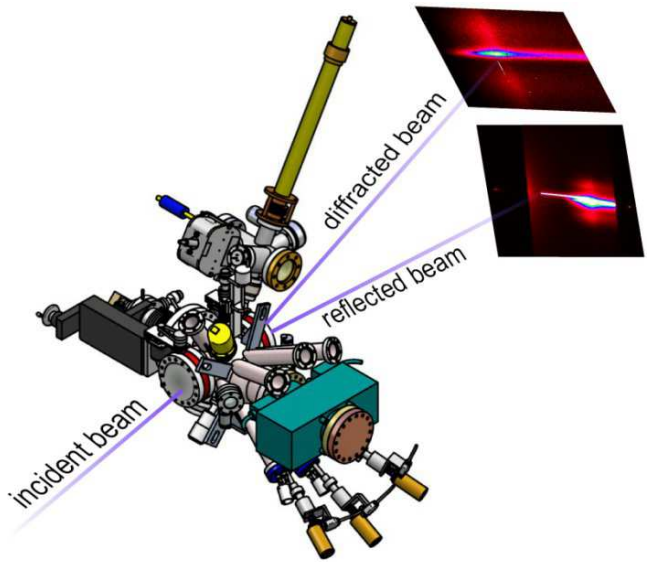}}

\caption{Sketch of the GID and GISAXS measurement geometry. \label{fig:gid gisaxs}}

\end{figure}

\begin{figure}

\label{GID}{\includegraphics[width=8.5cm]{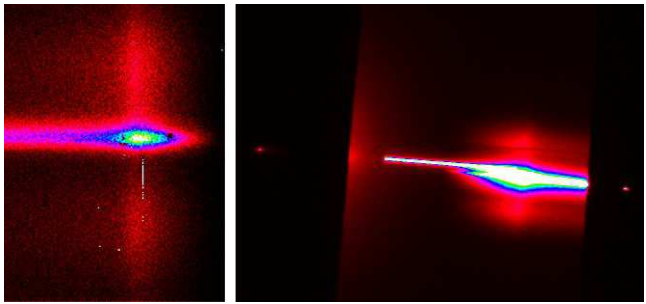}}

\caption{Left: GID signal of freestanding InAs QDs recorded with a MAXIPIX CCD
camera. Right: GISAXS simultaneously recorded with a MarCCD detector.\label{fig:gid gisaxs1}}

\end{figure}

\section{Exemplary results and discussion }

In this section, we present exemplary experimental results received
during \textit{in situ} measurements in GID geometry during post-growth annealing,
using the portable MBE system. We investigated a sample, grown at
the CFN, featuring InAs QD overgrown by a 1.3~ML AlAs layer and 5~nm GaAs. The X-ray measurements were performed at the surface diffraction
beamline ID03 at the ESRF. 

The sample was transported in the TC to the ESRF under UHV conditions.
Uninterrupted Power Supplies provided electricity for the pumping
system during the transport. An UHV connection between SC and TC was
established and the sample was inserted into the GC. The GC and SC
itself were then mounted on the diffractometers "Huber tower", where
a LASER sample alignment was done. 

A Maxipix single chip CCD detector was used to measure the scattered
intensities. The Maxipix gives an advantageous high time resolution,
the integration time was 1 second per frame. Furthermore, a 2D detector
enables us to measure a 3D volume of the reciprocal space. In GID configuration
an additional information about the scattering component
vertical to the surface is received. However, the CCD limits our resolution
in reciprocal space by detector size and detector-sample distance.

We chose the detector-sample distance to be d=130~cm. The Maxipix
chip has a side length of 14~mm which results in a resolution element
of $\pm0.34$~nm$^{-1}$ in reciprocal space. Thus we focused on the GaAs
bragg peak at $Q_{110}=31.416$~nm$^{-1}$ recording the diffuse intensity
at higher Q-values, neglecting the contribution of tensile strained
material. Due to the time constraint during the annealing process,
a line detector with analyzer crystal could not be applied.

\begin{figure}
\label{Before annealing}{\includegraphics[width=8.5cm]{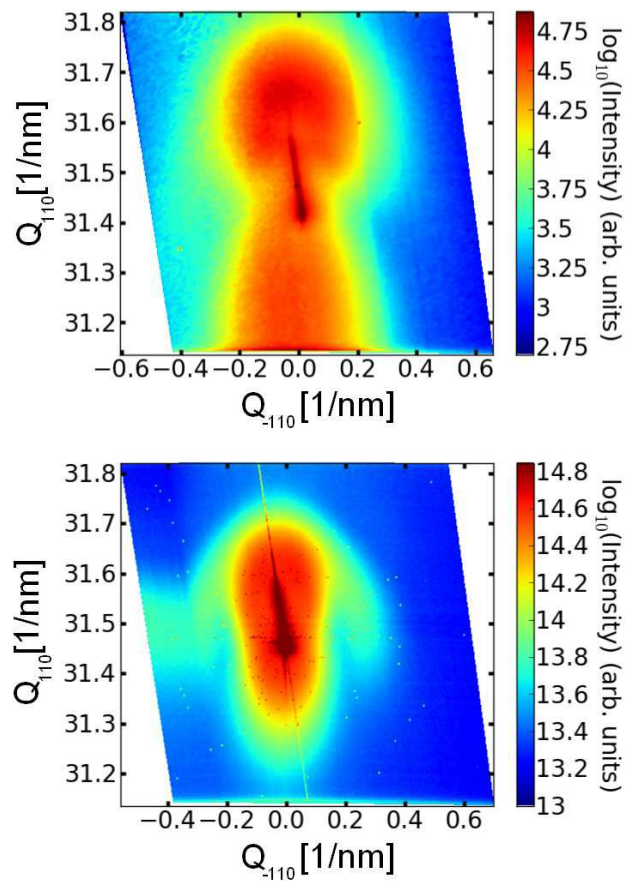}}

\caption{Reciprocal space maps of the {[}220{]} GaAs bragg peak measured \textit{in situ}
using a photon energy of $E=13$~keV at an incidence angle of $\alpha_{i}=0.5\text{\textdegree{}}$.
Top: before annealing. Bottom: after annealing (see text).\label{fig:RSM}}
\end{figure}

RSM of a {[}220{]} reflex were measured before and after annealing.
Figure \ref{fig:RSM} shows a RSM of the {[}220{]} GaAs reflection
before and after the annealing process. As it can be seen, the diffuse
scattering at higher and lower scattering vectors is reduced compared
to the state before the annealing process. This implies that the elastic
strain, induced by the lattice mismatch of the InAs/GaAs material
system has been partially relaxed due to material interdiffusion.
\cite{hu_morphology_2010,qiu_study_2010}

\begin{figure}
\label{Profiles along $Q_{-110}$ }{\includegraphics[width=8.5cm]{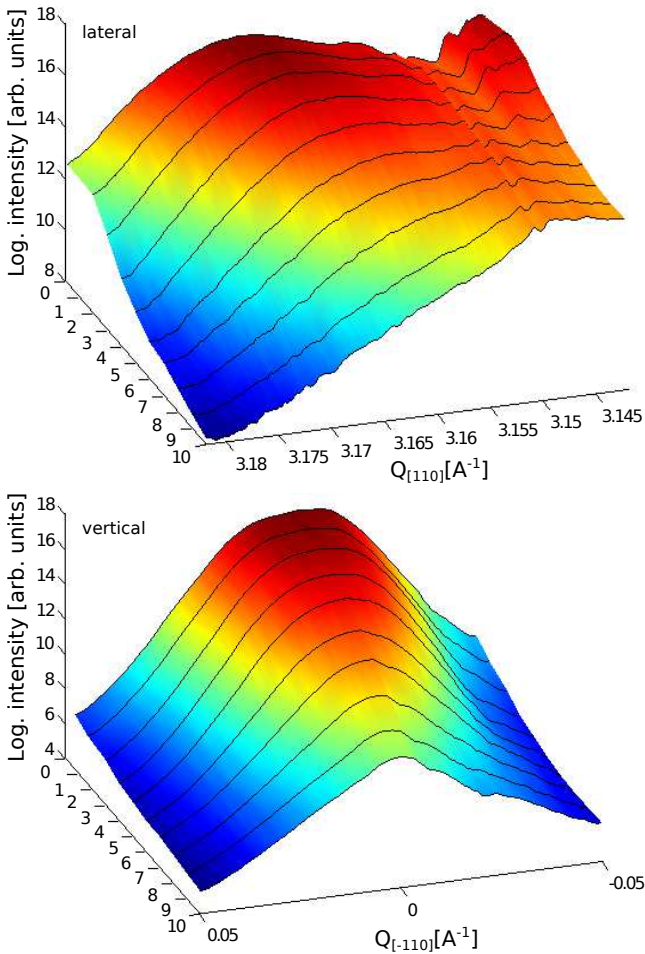}}

\caption{Top: Time evolution of scattered signal intensity profiles in $Q_{110}$ direction taken at $Q_{-110}=-0.0014$~1/A.
Bottom: Time evolution of profiles in $Q_{-110}$ taken at around $Q_{110}=3.1622$~1/A. The profiles can be understood as lateral and vertical slices through Fig.~\ref{fig:RSM}. The numbers indicate evolution of the measured signal during the annealing process as described in text. \label{fig:profiles}}
\end{figure}

A total of 27~RSMs have been measured \textit{in situ} during the annealing
process. The sample was annealed for 2~hours increasing the temperature
to 450\textdegree{}C stepwise. To monitor the evolution of the strained
GaAs substrate during annealing we show cuts through the RSM around
the compressed material peak at fixed $Q_{110}$ and $Q_{-110}$ positions
respectively for different annealing stages, see Fig.~\ref{fig:profiles}. The profiles are shown for the substrate temperatures of 90.3, 90.3, 188.0, 247.1, 292.3, 328.2, 359.8, 360.5, 361.3, 362.3~\textdegree{}C and annealing times of 4.5, 9.0, 36.0, 49.5, 63.0, 72.0, 81.0, 85.5, 90.0, 99.0 minutes.

As it can be seen from the $Q_{-110}$ profile plot, the intensity,
originating from the compressively strained GaAs substrate decreases
during the temperature ramp. The $Q_{110}$ profile plot indicates,
that this decrease is accompanied by a shift of the maximum position
from $Q_{110}=31.7~\textrm{nm}^{-1}$ to $Q_{110}=31.6~\textrm{nm}^{-1}$
before the maximum is vanishing. This process can be related to strain
relaxation due to material interdiffusion. 

On the other hand, indium desorption is competing with the diffusion process. This leads
to strain relaxation by gallium atoms substituting the desorbed
indium atoms. As a result, the indium-content of the QD decreases
\cite{jackson_monitoring_1997}. However, this process is known to
start at higher temperatures than the temperatures applied during
our annealing process \cite{heyn_stability_2002}. Thus, we assume that, the evolution of the scattered diffuse intensity is
largely caused the diffusion processes between the QDs and the surrounding media.

\section{Summary and Conclusions}

In this paper we present a new portable MBE system for \textit{in situ} X-ray
studies of III/V nanostructures. The system is suitable for investigations
at various synchrotron beamlines. The system's main features were
described and possible X-ray measurement methods such as GID and GISAXS
were found to be compatible to the geometrical constraints imposed
by the beryllium windows. 

The introduced geometries were tested and
first results obtained from \textit{in situ} XRD investigations of the annealing
process of InAs QDs were shown. The results indicate that the evolution
of the strain field induced by the QDs in the substrate can be successfully
investigated during the annealing process. 

We believe that the combination
of a portable MBE system with a synchrotron light source of high flux
and brilliance will help understanding dynamical processes during
the growth and post-growth processing of nanostructures, and offering
access to large variety of X-ray investigation methods at differently
specialized synchrotron beamlines. 

\section*{acknowledgments}
Authors would like to thank BMBF for financial support. The in situ experiments were performed on the ID03 beamline at the European Synchrotron Radiation Facility (ESRF), Grenoble, France. We are grateful to R. Felici and T. Dufrane at ESRF for providing assistance in using the beamline and dedicated support during the experiments. We also thank to the company "Createc" for implementation of the MBE system design.

\end{document}